# Critical noncolorings of the 600-cell proving the Bell-Kochen-Specker theorem

Mordecai Waegell* and P.K.Aravind**
Physics Department
Worcester Polytechnic Institute
Worcester, MA 01609
*caiw@wpi.edu ,**paravind@wpi.edu

ABSTRACT

Aravind and Lee-Elkin (1998) gave a proof of the Bell-Kochen-Specker theorem by showing that it is impossible to color the 60 directions from the center of a 600-cell to its vertices in a certain way. This paper refines that result by showing that the 60 directions contain many subsets of 36 and 30 directions that cannot be similarly colored, and so provide more economical demonstrations of the theorem. Further, these subsets are shown to be *critical* in the sense that deleting even a single direction from any of them causes the proof to fail. The critical sets of size 36 and 30 are shown to belong to orbits of 200 and 240 members, respectively, under the symmetries of the polytope. A comparison is made between these critical sets and other such sets in four dimensions, and the significance of these results is discussed.

## 1. Introduction

Some time back Lee-Elkin and one of us [1] gave a proof of the Bell-Kochen-Specker (BKS) theorem [2,3] by showing that it is impossible to color the 60 directions from the center of a 600-cell to its vertices in a certain way. This paper refines that result in two ways. Firstly, it shows that the 60 directions contain many subsets of 36 and 30 directions that cannot be similarly colored, and so provide more economical demonstrations of the theorem.  And, secondly, it shows that these subsets are *critical* in the sense that deleting even a single direction from any of them causes the proof to fail. The symmetries of the 600-cell, which had been ignored in our earlier work [1], are exploited in an essential way in arriving at these new results. Several critical sets of directions (or rays) in four dimensions are already known. Peres [4] proposed a set of 24 rays based on the geometry of a hypercube that possesses critical subsets of 18 and 20 rays [5-7]. Zimba and Penrose [8] proposed a set of 40 rays based on the geometry of a dodecahedron that possesses critical subsets of 28 rays. The present work adds two new critical sets to that list and, in the case of the new 30-ray sets, gives a "parity" proof of the BKS theorem that is as simple as the 18-ray proof of Cabello et.al. [5].

This paper is structured as follows. Section 2 reviews the BKS theorem and the non-coloring argument used to prove it. Section 3 introduces the 600-cell and describes all the geometrical facts about it needed in the present work. It then develops criteria for picking out subsets of its 60 directions, consisting of 36 directions in one case and 30 in another, that prove the BKS theorem. The criticality of these subsets is established, and the number of subsets of each type is also determined. Section 4 compares the present critical sets with the other such sets in four dimensions, and also discusses some applications of this work.

## 2. The Bell-Kochen-Specker (BKS) theorem

The Bell-Kochen-Specker theorem rules out the existence of deterministic noncontextual hidden variables theories. A proof of the theorem in a Hilbert space of dimension $d \geq 3$ can be given by exhibiting a finite set of rays [9] that cannot each be assigned the value 0 or 1 in such a way that (i) no two orthogonal rays are both assigned the value 1, and (ii) not all members of a set of $d$ mutually orthogonal rays are assigned the value 0. Bell [2] proved the theorem using a continuum argument, but Kochen and Specker [3] later exhibited a set of 117 directions in ordinary three-dimensional space that proved it. From a mathematical point of view, the BKS theorem is a corollary of a more powerful theorem proved earlier by Gleason [10].

Since the original work of Bell and of Kochen and Specker, numerous other proofs of the BKS theorem have been given in three [4], four [4-8] and higher [11,12,13] dimensions. Of particular interest for this work are the two other proofs of the theorem that have been given in four dimensions based on regular geometrical figures [4-8].

We adopt the language of Peres and speak of coloring rays red or green instead of labeling them 0 or 1, respectively. A proof of the BKS theorem based on a given set of rays then requires showing that it is impossible to color the rays red or green in such a way that no two orthogonal rays are both colored green and not all members of a complete orthogonal set are colored red. The only information needed for the proof is knowledge of the orthogonalities between the rays. This is sometimes summarized in the Kochen-Specker diagram, a graph whose vertices represent the rays and whose edges join vertices corresponding to orthogonal rays. For the 600-cell, the rays form a number of complete bases (i.e. sets of four mutually orthogonal rays) that include all orthogonalities between rays among them. This makes it possible to replace the Kochen-Specker diagram of the rays by their "basis table", and that is what we will do in this paper.

## 3. The 600-cell [14] and the BKS theorem

### 3.1 The 60 rays of the 600-cell

The 600-cell is a regular polytope in four-dimensional Euclidean space with 120 vertices distributed symmetrically on the surface of a four-dimensional sphere. The vertices come in antipodal pairs, with the members of each pair being diametrically opposite each other from the center of the sphere. Table 1 shows the coordinates of 60 of the vertices, with only one member taken from each antipodal pair. The center of the polytope is at the origin, so the coordinates of the omitted vertices are just the negatives of the ones shown. The directions from the origin to the 60 vertices shown define a system of 60 rays in a real four dimensional projective Hilbert space. It is this system of rays that was used in [1] to prove the BKS theorem, and that also serves as the basis for the present work.

The 600-cell is a remarkably symmetrical figure, whose symmetry group (of 14,400 elements) is transitive on all its vertices. What this means is that there is a symmetry operation that takes any vertex into any other vertex while leaving the figure as a whole invariant. The transitivity is not confined to the vertices but also extends to groups of vertices, such as bases and lines (to be defined below), and we will exploit this fact below.

Because it is a complicated figure, the structure of the 600-cell is not easy to grasp even when viewed in any of its two- or three-dimensional projections. One way of getting some feeling for it is to see how it can be decomposed into familiar three-dimensional figures. Let us look at the first, second, third and fourth nearest neighbors of the vertex labeled 1 in Table 1. The nearest neighbors, of which there are 12, are all the vertices whose first coordinate is $\tau$, and they lie at the vertices of an icosahedron. The second nearest neighbors, of which there are 20, are the vertices whose first coordinate is 1, and they lie at the vertices of a dodecahedron. The third nearest neighbors, of which there are 12, are the vertices whose first coordinate is $\kappa$, and they lie at the vertices of a second icosahedron. The fourth nearest neighbors, of which there are 15, are the vertices whose first coordinate is 0, and they make up half the vertices of an icosidodecahedron whose remaining vertices are the antipodes of the ones identified. The fifth, sixth and seventh nearest neighbors are repeats of the third, second and first, respectively, and the lone eighth nearest neighbor is the vertex antipodal to 1. Because of the transitivity of the polytope on its vertices, each vertex has an identical pattern of neighbors.

A still more important division of the 600-cell, for our purposes, is into the simpler four-dimensional figures of which it is made up. Consider the 12 vertices in each of the separate blocks of Table 1. The four vertices in any row of a block are in mutually orthogonal directions from the center of the 600-cell and, together with their antipodes, make up the vertices of a figure known as a cross polytope or 16-cell. The eight vertices in any pair of rows of a block, together with their antipodes, make up the vertices of tesseract or 8-cell. And all 12 vertices in a block, together with their antipodes, make up the vertices of a 24-cell, which is so called because its bounding cells are 24 octahedra. The blocks in Table 1 illustrate the fact that the vertices of a 600-cell can be partitioned into those of five disjoint 24-cells that have no vertices in common.

If the entries in Table 1 are interpreted as rays in Hilbert space, then the four rays in any row form a basis of four mutually orthogonal rays. Table 1 can then be viewed as a listing of 15 of the bases formed by the 60 rays of the 600-cell. The three (mutually unbiased) bases in any block are associated with one of the 24-cells into which the 600-cell can be decomposed.

3.2 The 75 bases of the 600-cell

A 600-cell can be partitioned into disjoint 24-cells in ten different ways. These ten ways are shown in Table 2, in which each horizontal row, as well as each vertical column, of blocks represents such a partition. The table also demonstrates that a 600-cell has 25 distinct 24-cells inscribed in it, with five 24-cells meeting at each vertex. Since each 24-cell gives rise to three bases, the 60 rays of the 600-cell give rise to 75 bases altogether. It can be verified by direct computation that there are no bases other than these and also no orthogonalities between rays not already contained in these bases. So Table 2 (the "basis" table) is indeed a complete substitute for the Kochen-Specker diagram of these rays and the only input needed for a proof of the BKS theorem based on them.

The 60 rays and 75 bases of the 600-cell form a configuration in which each ray occurs in 5 bases and each basis contains 4 rays. This fact is captured in the equality 75 x 4 = 60 x 5, which says that summing the rays over the bases yields each ray exactly five times in the total count.

Each ray is orthogonal to exactly 15 others, and these are just its companions in the five bases in which it occurs. The transitivity of the rays is readily apparent in the basis table.

For reference purposes we attach a pair of labels (an unprimed letter from A to E and a primed letter from A' to E') to each of the 24-cells in Table 2. The cells in a column share a common unprimed letter, while those in a row share a common primed letter.

3.3 Symmetries of the 600-cell

The symmetry operations of the 600-cell are (proper or improper) orthogonal transformations in four-dimensional Euclidean space. The transformations can be characterized by their effect on a set of vertices, such as 1,2,3 and 4, that are in mutually orthogonal directions from the center of the polytope. The symmetry operations consist of all mappings of the fiducial set 1,2,3,4 into the 75 sets shown in Table 2, but with the following qualifications: (i) the set 1,2,3,4 can be mapped into only the even permutations of the sets x,y,z,w shown in Table 2; and (ii) in any mapping allowed under (i), such as 1 $\rightarrow$ y, 2 $\rightarrow$ x, 3 $\rightarrow$ w, 4 $\rightarrow$ z, either a positive or negative sign can be affixed in front of each of x,y,z and w. These conditions fix the total number of symmetry operations as 75 x 12 x $2^4$ = 14,400, the number stated earlier. In the projective Hilbert space of rays, where there is no distinction between a direction and its opposite, the number of symmetry operations goes down by a factor of two to 7200.

The symmetries of the 600-cell have been exploited in setting up Table 2. The rows of the table are cycled into each other by the period-5 rotation $|13\rangle\langle 1|+|14\rangle\langle 2|+|15\rangle\langle 3|+|16\rangle\langle 4|$ and the columns by the period-5 rotation $|31\rangle\langle 1|+|42\rangle\langle 2|-|51\rangle\langle 3|+|16\rangle\langle 4|$, where the bras and kets are row and column vectors representing the rays. An alternative way of describing these rotations is by specifying the permutations they perform on the rays as one goes from one row (or column) to the next. The permutations are easily picked out by comparing neighboring rows or columns. For either the rows or the columns, the permutation can be written as a product of 12 cycles of length 5 each. The permutation that cycles the rows into each other is even more simply described: simply add 12, modulo 60, to the numbers in any row to generate the numbers in the row below it.

3.4 The 24-cell: points, lines and Reye's configuration

Any 24-cell can be interpreted as a "Reye's configuration" [15], which is a set of 12 "points" and 16 "lines" with the property that three "points" lie on every "line" and four "lines" pass through every "point". The quotes have been put around "points" and "lines" to indicate that these entities need not literally be geometrical points or lines but can be any entities that satisfy the stated incidence relations. A 24-cell can be interpreted as a Reye's configuration if each of its 12 rays is regarded as a "point" and any three rays as a "line" if the components of one can be expressed as a linear combination of those of the other two. To verify the truth of this assertion, consider the rays 1-12 associated with the first 24-cell of Table 1. It is straightforward to check that these rays, interpreted as "points", make up the 16 "lines" given by the triples {1,5,9}, {1,6,10}, {1,7,11}, {1,8,12}, {2,5,10}, {2,6,9}, {2,7,12}, {2,8,11}, {3,5,11}, {3,6,12}, {3,7,9}, {3,8,10}, {4,5,12}, {4,6,11}, {4,7,10} and {4,8,9}. Table 3 gives a simple prescription for generating the lines associated with any of the 24-cells in Table 2.

Since the 600-cell contains 25 different 24-cells within it, and each 24-cell gives rise to 16 lines, it might be thought that the 600-cell has 400 lines in it altogether. However it turns out that each line is shared by two 24-cells, so there are just 200 unique lines in all. The lines also come in *dual pairs*, with all the rays on any line being orthogonal to all the rays on its dual. The 200 lines thus break up into 100 dual pairs. A complete listing of the lines, by dual pairs, is shown in Table 4. The letters labeling the rows and the columns help to identify the 24-cells in which each of the lines of the dual pair originates. Table 4 shares in common with Table 2 the feature that each of its rows and columns of blocks contains the numbers from 1 to 60 exactly once.

3.5 Critical sets of 30 rays

These critical sets are obtained by keeping only the rays belonging to one of the chains of dual line pairs shown in Table 5. On applying the procedure explained in the caption of Table 5 to Table 4, one can pick out all the 30-ray critical sets. There are 12 such sets for each row and column of Table 4, and therefore 240 such sets altogether.

It remains to show that these sets are indeed critical. This involves showing, firstly, that any such set cannot be properly colored and, secondly, that it becomes colorable if even a single ray is deleted from it. It is sufficient to demonstrate these properties for any one of the 30-ray sets, since the same conclusions then follow for all the other sets from symmetry.

Consider the 30-ray set obtained by applying the chain AB-BC-CD-DE-EA to the first row of Table 4. The 30 rays so obtained form 15 complete bases, which are shown in bold in Table 6. The fact that each ray occurs twice in these bases suffices to give a parity proof of the BKS theorem. To see why, recall, from Sec.2, that the BKS theorem can be proved using these 30 rays if it can be shown that they cannot each be colored red or green in such a way that the 15 bases formed only of them each contain one green ray and three red rays. However the impossibility of this task is easily seen because, on the one hand, there must be an odd number of green rays in these bases (since there must be exactly one green ray per basis) but, on the other hand, there must also be an even number of green rays (since each green ray occurs twice).

A proof of criticality is easily given if one uses the fact that the 30-ray set has a symmetry group that is transitive on all its members (just like the original 60-ray set). Because of this, it suffices to delete any one ray and show that the rest then become colorable. To this end, we note that deleting ray 3 and coloring rays 25, 45, 55, 16, 4, 32, and 49 green and all the other rays red achieves a satisfactory coloring of all bases (including partially deleted ones) involving any of the 29 surviving rays.

An interesting feature of the 30-ray sets is that they come in complementary pairs having no rays in common (and therefore spanning all 60 rays when taken together). A pair of such sets is shown in Table 6, with the 15 bases made up by the rays of the second set being shown in italics. There are 120 such complementary pairs of 30-ray sets within this 60-ray system.

3.6 Critical sets of 36 rays

These critical sets are obtained by keeping only the rays belonging to one of the chains of dual line pairs shown in Table 7. The procedure for constructing these sets is similar to that for the 30-ray sets. Each row and column of Table 4 yields 10 critical sets of this type, and so there are 200 such critical sets altogether.

As before, it suffices to pick a single 36-ray set and show that it is both uncolorable and critical, and symmetry will then guarantee the result for the other sets. We pick the set obtained by applying the first chain in Table 7 to the first row of Table 4. The 36 rays thus selected are shown in bold in Table 8. Leaving aside the 9 bases in which none of these rays occur, the remaining bases are of three types: (i) 18 "intact" bases involving only the 36 surviving rays, (ii) 12 "almost intact" bases, each made up of three surviving rays and one deleted ray, and (iii) 36 "half intact" bases, each made up of two surviving rays and two deleted rays. Unlike the case of the 30-ray sets, no parity proof of the BKS theorem is possible in this case. However Table 9 gives a straightforward "proof-tree" argument demonstrating the uncolorability of these rays. This argument makes use of only the 30 "intact" and "almost intact" bases, and ignores the "half intact" bases. The reason for ignoring the latter bases was to make the proof rely on as few orthogonalities (between the surviving rays) as possible.

An alternative proof of the uncolorability of the 36-ray set, that again makes use of only the 30 bases mentioned above, can be given by noting that the 18 "intact" bases can be satisfactorily colored in a total of 448 different ways, but that in each of these cases at least one "half intact" basis has more than one green ray in it. We wrote a computer program that allowed us to arrive at this conclusion.

Since the 36-ray set (like the 30-ray set) is transitive on all its rays, its criticality can be established by deleting any one ray and showing that the remainder can be properly colored. If one deletes ray 1, and colors rays 11, 4, 22, 54, 18, 5, 35, and 25 green and all the others red, it is easy to check that all the 66 bases involving any of the 35 surviving rays are properly colored.

**4. Discussion**

We have shown that the system of 60 rays derived from the vertices of a 600-cell possesses 30-ray and 36-ray critical sets. Although we think it highly unlikely that smaller critical sets exist, we do not have a proof of this conjecture. Either a proof of the conjecture or a counterexample to it would be of interest. A recent paper by Pavicic et al [29] uses lattice theory and MMP diagrams to determine all Kochen-Specker sets of rays that can be obtained as subsets of 24 rays with components from the set $\{-1,0,1\}$. It is possible that an extension of these methods might allow a resolution of our conjecture.

Table 10 lists the smallest known critical sets of rays in four dimensions. They all occur as subsets of one of three systems of rays: (i) the 24 rays of Peres [4], derived from a pair of dual 24-cells, (ii) the 40 rays of Zimba and Penrose [8], derived from the geometry of a dodecahedron, and (iii) the 60 rays of the present work, derived from the vertices of a 600-cell.

The smallest of these sets are the 18-ray sets, which occur as subsets of the 24 rays of Peres. They were originally discovered through a computer search [5], but later a simple construction was found for them [7]: they can be obtained by omitting all rays lying on a dual pair of lines, one from each of the dual 24-cells defining the system. Because each 24-cell has 16 lines in it, and each line in a cell has a unique dual in the other, there are exactly 16 such critical sets. It was this simple observation that motivated the present work. Since a 600-cell contains a large number of 24-cells within it, we felt that the strategy of cancelling dual pairs of lines from different 24-cells might again prove useful in leading to critical sets. This indeed turned out to be the case, but a couple of differences between the 600-cell and the Peres system should be kept in mind. Firstly, the 600-cell does not contain the dual of any of its 24-cells, and so does not contain the Peres rays as a subset (else its noncolorability would be trivial). And, secondly, the critical sets result not from the cancellation of a single dual pair of lines but four pairs (for the 36-ray sets) or five pairs (for the 30-ray sets).

The Zimba-Penrose system is different from the other two systems in a couple of ways. Firstly, its 40 rays are essentially complex, whereas the rays in the other two systems can be chosen to be purely real. And, secondly, Reye's configuration does not make its appearance in this system, although it does have several other point-line configurations of interest, as noted by Zimba and Penrose [8]. Zimba and Penrose pointed out how the five 28-ray critical sets could be extracted by using the fact that there are five ways of inscribing a cube in a dodecahedron. We recently discovered how to obtain the 32-ray critical sets by cancelling dual pair of lines, although each line now consists of four points [16].

The main interest of the new results presented here stems from the expanded view they give of the tension between four dimensional geometry and quantum noncontextuality. Although three dimensions are sufficient to rule out noncontextuality, ascending to one higher dimension makes the task much simpler. This is most strikingly evident from the fact that parity proofs of the BKS theorem first become possible in four dimensions. The first, second and fifth critical sets in Table 10 yield three such proofs, and they are much simpler than any of the three dimensional proofs (although it should be added, in fairness, that the importance and geometric interest of finding purely three-dimensional proofs is undeniable). It is also worth remarking that while all the more recent three dimensional proofs make use of cubic symmetry, the four dimensional proofs exploit different symmetries (namely, those of the icosahedron and enlargements of the 24-cell). The way in which dimensionality and symmetry influence the existence of critical sets is a theme that has been only partially explored to date [13,17,18], and it is likely that there are some interesting results that remain to be discovered here.

We have mentioned only the BKS theorem in this paper. However it should be added that the present proof, like all proofs of the BKS theorem in any dimension, can be extended to a proof of Bell's nonlocality theorem by making use of the right sort of entanglement [19].

Recently there has been considerable interest in experimental tests of quantum noncontextuality [20]-[25]. Cabello [20] and Badziag et.al.[21] have shown how to derive a variety of inequalities that are satisfied by noncontextual theories but violated by all quantum states, even ones thought to have a totally classical character. Experiments with ions [24] and photons [25] have shown a violation of some of these inequalities, thereby casting a new light on the nonclassicality of even unentangled quantum states.

The BKS proof of this paper can be used to derive several further inequalities of the type proposed in [20] and [21]. The interest of this exercise would be to see if the gap between noncontextual theories and quantum mechanics can be made wider than in any of the cases studied hitherto. Using a 30-ray critical set of the 600-cell, together with the 15 complete bases in which these rays occur, one can construct an inequality similar to Eq.(1) in [20]. The upper bound of this inequality for noncontextual theories is 13, but quantum mechanics predicts that the inequality is actually an equality with a value of 15 (independent of the state considered). Thus there is again a difference of 2 between the two predictions, as was the case with the 18 vectors and 9 bases treated by Cabello [20]. However [20] points out that if one chooses all the 24 rays of Peres, along with the 24 bases formed by them, the gap between noncontextual theories and quantum mechanics becomes wider. This is attributed to the fact that the 24-ray set is not critical but contains a large number of critical subsets within it. In the same vein, it is possible that if one uses all 60 rays and 75 bases of the 600-cell to construct an inequality, the gap between noncontextual realism and quantum mechanics can be made even larger because of the large number of critical subsets involved in the present case. An experimental realization of this configuration may not be easy but it may still be interesting, from a theoretical point of view, to see how hard one can push on geometry to drive a wedge between noncontextual realism and the quantum theory.

BKS proofs are also of interest in connection with quantum "pseudo-telepathy" [26-28], two-party tasks that cannot be completed with the aid of classical resources alone but require the use of entanglement. The present proof provides a new instance of such a task. Whether there are any further applications of the new critical sets found here remains to be seen.

# THE 60 RAYS OF THE 600-CELL

| | | | |
|---|---|---|---|
| $1 = 2000$ | $2 = 0200$ | $3 = 0020$ | $4 = 0002$ |
| $5 = 1111$ | $6 = 11\bar{1}\bar{1}$ | $7 = 1\bar{1}1\bar{1}$ | $8 = 1\bar{1}\bar{1}1$ |
| $9 = 1\overline{111}$ | $10 = 1\bar{1}11$ | $11 = 11\bar{1}1$ | $12 = 111\bar{1}$ |
| $13 = \kappa 0\tau\bar{1}$ | $14 = 0\kappa 1\bar{\tau}$ | $15 = \tau\bar{1}\kappa 0$ | $16 = 1\tau 0\kappa$ |
| $17 = \tau\kappa 0\bar{1}$ | $18 = 10\kappa\tau$ | $19 = \kappa\bar{\tau}10$ | $20 = 01\tau\kappa$ |
| $21 = 1\kappa\tau 0$ | $22 = \tau 0\bar{1}\kappa$ | $23 = 0\tau\bar{\kappa}1$ | $24 = \kappa\bar{1}0\bar{\tau}$ |
| $25 = \tau 01\kappa$ | $26 = 0\tau\bar{\kappa}1$ | $27 = 1\bar{\kappa}\tau 0$ | $28 = \kappa 10\bar{\tau}$ |
| $29 = 0\kappa 1\tau$ | $30 = \tau 1\bar{\kappa}0$ | $31 = \kappa 0\tau\bar{1}$ | $32 = 1\bar{\tau}0\kappa$ |
| $33 = \tau\bar{\kappa}0\bar{1}$ | $34 = 01\bar{\tau}\kappa$ | $35 = 10\bar{\kappa}\tau$ | $36 = \kappa\tau 10$ |
| $37 = \tau 0\bar{1}\bar{\kappa}$ | $38 = 0\tau\kappa\bar{1}$ | $39 = 1\bar{\kappa}\tau 0$ | $40 = \kappa 10\tau$ |
| $41 = \tau 1\kappa 0$ | $42 = 0\kappa\bar{1}\bar{\tau}$ | $43 = 1\tau 0\bar{\kappa}$ | $44 = \kappa 0\bar{\tau}1$ |
| $45 = 01\tau\kappa$ | $46 = \tau\bar{\kappa}01$ | $47 = \kappa\tau\bar{1}0$ | $48 = 10\kappa\bar{\tau}$ |
| $49 = \kappa 0\tau 1$ | $50 = 0\kappa\bar{1}\tau$ | $51 = \tau\bar{1}\bar{\kappa}0$ | $52 = 1\tau 0\bar{\kappa}$ |
| $53 = 10\bar{\kappa}\bar{\tau}$ | $54 = \tau\kappa 01$ | $55 = 01\tau\bar{\kappa}$ | $56 = \kappa\bar{\tau}10$ |
| $57 = \tau 01\bar{\kappa}$ | $58 = 1\kappa\bar{\tau}0$ | $59 = \kappa\bar{1}0\tau$ | $60 = 0\tau\kappa 1$ |

Table 1. Coordinates of 60 of the vertices of a 600-cell, with $\tau = (\sqrt{5}+1)/2$, $\kappa = 1/\tau$, a bar over a number indicating its negative, and commas omitted between coordinates. The vertices are chosen to lie on a sphere of radius 2 centered at the origin. Each entry can also interpreted as a ray in a real four-dimensional Hilbert space, with the numbers representing its components in an orthonormal basis. The rays in any row of the table constitute a basis, and the entire table shows 15 of the bases formed by the 60 rays of the 600-cell. The twelve vectors in any block (together with their antipodes) make up the vertices of a 24-cell, and the table shows how the vertices of a 600-cell can be partitioned into those of five disjoint 24-cells.

# THE 75 BASES OF THE 600-CELL

|     | A | B | C | D | E |
|-----|---|---|---|---|---|
| A'  | 1  2  3  4<br>5  6  7  8<br>9  10  11  12 | 31  42  51  16<br>38  24  58  25<br>56  45  17  35 | 22  60  39  28<br>18  47  33  55<br>13  32  50  41 | 57  23  27  40<br>36  53  20  46<br>43  49  30  14 | 44  29  15  52<br>59  26  37  21<br>34  19  48  54 |
| B'  | 13  14  15  16<br>17  18  19  20<br>21  22  23  24 | 43  54  3  28<br>50  36  10  37<br>8  57  29  47 | 34  12  51  40<br>30  59  45  7<br>25  44  2  53 | 9  35  39  52<br>48  5  32  58<br>55  1  42  26 | 56  41  27  4<br>11  38  49  33<br>46  31  60  6 |
| C'  | 25  26  27  28<br>29  30  31  32<br>33  34  35  36 | 55  6  15  40<br>2  48  22  49<br>20  9  41  59 | 46  24  3  52<br>42  11  57  19<br>37  56  14  5 | 21  47  51  4<br>60  17  44  10<br>7  13  54  38 | 8  53  39  16<br>23  50  1  45<br>58  43  12  18 |
| D'  | 37  38  39  40<br>41  42  43  44<br>45  46  47  48 | 7  18  27  52<br>14  60  34  1<br>32  21  53  11 | 58  36  15  4<br>54  23  9  31<br>49  8  26  17 | 33  59  3  16<br>12  29  56  22<br>19  25  6  50 | 20  5  51  28<br>35  2  13  57<br>10  55  24  30 |
| E'  | 49  50  51  52<br>53  54  55  56<br>57  58  59  60 | 19  30  39  4<br>26  12  46  13<br>44  33  5  23 | 10  48  27  16<br>6  35  21  43<br>1  20  38  29 | 45  11  15  28<br>24  41  8  34<br>31  37  18  2 | 32  17  3  40<br>47  14  25  9<br>22  7  36  42 |

Table 2. The 75 bases formed by the 60 rays of the 600-cell (with the rays numbered as in Table 1). Each block of 12 rays represents a 24-cell, with its three rows representing three mutually unbiased bases. The first column of the table is essentially a repeat of Table 1, while the next four columns show four alternative decompositions of the 600-cell into five disjoint 24-cells. The rows show five additional decompositions of the same kind, bringing the total number of such decompositions to ten. There are 25 distinct 24-cells inscribed in the 600-cell and they give rise to 25 x 3 = 75 bases altogether. The letters at the head of each row and column serve to label the 24-cells, with each 24-cell being labeled by an unprimed and a primed letter. Any two 24-cells have either no rays in common (such as cells AA' and AB') or three rays in common (such as cells AA' and DB'). Each ray occurs in five 24-cells and five bases, and its fifteen companions in the bases in which it occurs are the only other rays it is orthogonal to. Thus this basis table is completely equivalent to the Kochen-Specker diagram of this set of rays.

## LINES OF THE 24-CELL

| a | b | c | d |
|---|---|---|---|
| e | f | g | h |
| i | j | k | l |

$a,e,i$  $b,e,j$  $c,e,k$  $d,e,l$
$a,f,j$  $b,f,i$  $c,f,l$  $d,f,k$
$a,g,k$  $b,g,l$  $c,g,i$  $d,g,j$
$a,h,l$  $b,h,k$  $c,h,j$  $d,h,i$

Table 3. The box at the top shows the 12 rays $a,b,...,l$ of a 24-cell, with each row representing a basis. Shown below are the 16 "lines" of the 24-cell, with each line being indicated by the three rays (or "points") on it. This template can be used to generate the 16 lines associated with each of the 24-cells in Table 2, and this has been done in Table 4.

# LINES OF THE 600-CELL

|   |   | A<br>B |    | A<br>C |    | A<br>D |    | A<br>E |    | B<br>C |    | B<br>D |    | B<br>E |    | C<br>D |    | C<br>E |    | D<br>E |    |
|---|---|----|----|----|----|----|----|----|----|----|----|----|----|----|----|----|----|----|----|----|----|
| A´ | B´ | 3  | 16 | 2  | 13 | 1  | 14 | 4  | 15 | 25 | 28 | 35 | 36 | 31 | 29 | 32 | 30 | 33 | 34 | 27 | 26 |
|    |    | 8  | 17 | 7  | 18 | 5  | 20 | 6  | 19 | 45 | 47 | 42 | 43 | 38 | 37 | 39 | 40 | 41 | 44 | 46 | 48 |
|    |    | 10 | 24 | 12 | 22 | 9  | 23 | 11 | 21 | 51 | 50 | 58 | 57 | 56 | 54 | 55 | 53 | 60 | 59 | 49 | 52 |
| A´ | C´ | 2  | 25 | 3  | 28 | 4  | 27 | 1  | 26 | 24 | 22 | 17 | 20 | 16 | 15 | 13 | 14 | 18 | 19 | 23 | 21 |
|    |    | 6  | 31 | 5  | 32 | 7  | 30 | 8  | 29 | 42 | 41 | 38 | 40 | 45 | 48 | 47 | 46 | 39 | 37 | 43 | 44 |
|    |    | 9  | 35 | 11 | 33 | 10 | 36 | 12 | 34 | 56 | 55 | 51 | 49 | 58 | 59 | 60 | 57 | 50 | 52 | 53 | 54 |
| A´ | D´ | 1  | 38 | 4  | 39 | 3  | 40 | 2  | 37 | 17 | 18 | 16 | 14 | 24 | 21 | 22 | 23 | 13 | 15 | 20 | 19 |
|    |    | 7  | 42 | 8  | 41 | 6  | 43 | 5  | 44 | 31 | 32 | 25 | 27 | 35 | 34 | 33 | 36 | 28 | 26 | 30 | 29 |
|    |    | 11 | 45 | 9  | 47 | 12 | 46 | 10 | 48 | 58 | 60 | 56 | 53 | 51 | 52 | 50 | 49 | 55 | 54 | 57 | 59 |
| A´ | E´ | 4  | 51 | 1  | 50 | 2  | 49 | 3  | 52 | 16 | 13 | 24 | 23 | 17 | 19 | 18 | 20 | 22 | 21 | 14 | 15 |
|    |    | 5  | 56 | 6  | 55 | 8  | 53 | 7  | 54 | 35 | 33 | 31 | 30 | 25 | 26 | 28 | 27 | 32 | 29 | 36 | 34 |
|    |    | 12 | 58 | 10 | 60 | 11 | 57 | 9  | 59 | 38 | 39 | 45 | 46 | 42 | 44 | 41 | 43 | 47 | 48 | 40 | 37 |
| B´ | C´ | 15 | 28 | 14 | 25 | 13 | 26 | 16 | 27 | 3  | 2  | 10 | 9  | 8  | 6  | 7  | 5  | 12 | 11 | 1  | 4  |
|    |    | 20 | 29 | 19 | 30 | 17 | 32 | 18 | 31 | 37 | 40 | 47 | 48 | 43 | 41 | 44 | 42 | 45 | 46 | 39 | 38 |
|    |    | 22 | 36 | 24 | 34 | 21 | 35 | 23 | 33 | 57 | 59 | 54 | 55 | 50 | 49 | 51 | 52 | 53 | 56 | 58 | 60 |
| B´ | D´ | 14 | 37 | 15 | 40 | 16 | 39 | 13 | 38 | 8  | 7  | 3  | 1  | 10 | 11 | 12 | 9  | 2  | 4  | 5  | 6  |
|    |    | 18 | 43 | 17 | 44 | 19 | 42 | 20 | 41 | 36 | 34 | 29 | 32 | 28 | 27 | 25 | 26 | 30 | 31 | 35 | 33 |
|    |    | 21 | 47 | 23 | 45 | 22 | 48 | 24 | 46 | 54 | 53 | 50 | 52 | 57 | 60 | 59 | 58 | 51 | 49 | 55 | 56 |
| B´ | E´ | 13 | 50 | 16 | 51 | 15 | 52 | 14 | 49 | 10 | 12 | 8  | 5  | 3  | 4  | 2  | 1  | 7  | 6  | 9  | 11 |
|    |    | 19 | 54 | 20 | 53 | 18 | 55 | 17 | 56 | 29 | 30 | 28 | 26 | 36 | 33 | 34 | 35 | 25 | 27 | 32 | 31 |
|    |    | 23 | 57 | 21 | 59 | 24 | 58 | 22 | 60 | 43 | 44 | 37 | 39 | 47 | 46 | 45 | 48 | 40 | 38 | 42 | 41 |
| C´ | D´ | 27 | 40 | 26 | 37 | 25 | 38 | 28 | 39 | 9  | 11 | 6  | 7  | 2  | 1  | 3  | 4  | 5  | 8  | 10 | 12 |
|    |    | 32 | 41 | 31 | 42 | 29 | 44 | 30 | 43 | 15 | 14 | 22 | 21 | 20 | 18 | 19 | 17 | 24 | 23 | 13 | 16 |
|    |    | 34 | 48 | 36 | 46 | 33 | 47 | 35 | 45 | 49 | 52 | 59 | 60 | 55 | 53 | 56 | 54 | 57 | 58 | 51 | 50 |
| C´ | E´ | 26 | 49 | 27 | 52 | 28 | 51 | 25 | 50 | 6  | 5  | 2  | 4  | 9  | 12 | 11 | 10 | 3  | 1  | 7  | 8  |
|    |    | 30 | 55 | 29 | 56 | 31 | 54 | 32 | 53 | 20 | 19 | 15 | 13 | 22 | 23 | 24 | 21 | 14 | 16 | 17 | 18 |
|    |    | 33 | 59 | 35 | 57 | 34 | 60 | 36 | 58 | 48 | 46 | 41 | 44 | 40 | 39 | 37 | 38 | 42 | 43 | 47 | 45 |
| D´ | E´ | 39 | 52 | 38 | 49 | 37 | 50 | 40 | 51 | 1  | 4  | 11 | 12 | 7  | 5  | 8  | 6  | 9  | 10 | 3  | 2  |
|    |    | 44 | 53 | 43 | 54 | 41 | 56 | 42 | 55 | 21 | 23 | 18 | 19 | 14 | 13 | 15 | 16 | 17 | 20 | 22 | 24 |
|    |    | 46 | 60 | 48 | 58 | 45 | 59 | 47 | 57 | 27 | 26 | 34 | 33 | 32 | 30 | 31 | 29 | 36 | 35 | 25 | 28 |

Table 4. The 200 lines of the 600-cell, with each box showing a dual pair (displayed vertically). The labels at the head of each row and column serve to identify the 24-cells in which the lines originate according to the following rule: the left line originates in the cells XU' and YV' and the right line in the cells XV' and YU' (where X and Y are the upper and lower column labels and U' and V' the left and right row labels). As an example, the line {3,8,10} originates in the cells AA' and BB' and its dual {16,17,24} in the cells AB' and BA'. Note that the lines in each column span all 60 rays of the 600-cell without any overlap, and that the same is true of the lines in the rows.

## PATTERNS OF 30-RAY CRITICAL SETS

      AB-BC-CD-DE-EA        AC-CE-EB-BD-DA
      AB-BD-DC-CE-EA        AC-CB-BE-ED-DA
      AB-BC-CE-ED-DA        AC-CD-DB-BE-EA
      AB-BE-EC-CD-DA        AC-CB-BD-DE-EA
      AB-BD-DE-EC-CA        AD-DC-CB-BE-EA
      AB-BE-ED-DC-CA        AD-DB-BC-CE-EA

TABLE 5. A 30-ray critical set can be obtained by keeping all the rays in any row of Table 4 belonging to one of the chains of dual line pairs shown above. A pair of letters, such as AB, indicates the dual line pair, in the selected row, lying in the column headed by those letters. For example, if the chain AB-BE-EC-CD-DA is applied to the third row, the label AB picks out the rays 1,7,11,38,42 and 45, BE picks out the rays 24,35,51,21,34 and 52, and so on. The chain AB-BC-CD-DE-EA applied to the first row yields the critical set shown in bold in Table 6. The two chains listed on the same line are complementary in the sense that they have no rays in common (and therefore span all 60 rays when taken together). If all the letters above are replaced by their primed counterparts and the chains are applied to the columns (in the same manner as the rows), the remaining 30-ray critical sets result.

**TWO COMPLEMENTARY 30-RAY CRITICAL SETS**

|     | A | B | C | D | E |
|-----|---|---|---|---|---|
| A'  | 1 2 3 4<br>5 6 7 8<br>9 10 11 12 | 31 42 51 16<br>38 24 58 25<br>56 45 17 35 | 22 60 39 28<br>18 47 33 55<br>13 32 50 41 | 57 23 27 40<br>36 53 20 46<br>43 49 30 14 | 44 29 15 52<br>59 26 37 21<br>34 19 48 54 |
| B'  | 13 14 15 16<br>17 18 19 20<br>21 22 23 24 | 43 54 3 28<br>50 36 10 37<br>8 57 29 47 | 34 12 51 40<br>30 59 45 7<br>25 44 2 53 | 9 35 39 52<br>48 5 32 58<br>55 1 42 26 | 56 41 27 4<br>11 38 49 33<br>46 31 60 6 |
| C'  | **25 26 27 28**<br>29 30 31 32<br>*33 34 35 36* | **55 6 15 40**<br>2 48 22 49<br>*20 9 41 59* | **46 24 3 52**<br>42 11 57 19<br>*37 56 14 5* | **21 47 51 4**<br>60 17 44 10<br>*7 13 54 38* | **8 53 39 16**<br>23 50 1 45<br>*58 43 12 18* |
| D'  | 37 38 39 40<br>*41 42 43 44*<br>**45 46 47 48** | 7 18 27 52<br>*14 60 34 1*<br>**32 21 53 11** | 58 36 15 4<br>*54 23 9 31*<br>**49 8 26 17** | 33 59 3 16<br>*12 29 56 22*<br>**19 25 6 50** | 20 5 51 28<br>*35 2 13 57*<br>**10 55 24 30** |
| E'  | **49 50 51 52**<br>53 54 55 56<br>*57 58 59 60* | **19 30 39 4**<br>26 12 46 13<br>*44 33 5 23* | **10 48 27 16**<br>6 35 21 43<br>*1 20 38 29* | **45 11 15 28**<br>24 41 8 34<br>*31 37 18 2* | **32 17 3 40**<br>47 14 25 9<br>*22 7 36 42* |

TABLE 6. Shown in bold are the 15 bases formed by the 30-ray critical set picked out by applying the chain AB-BC-CD-DE-EA of Table 5 to the first row of Table 4. Shown in italics are the 15 bases formed by the 30 rays of the complementary critical set obtained by applying the chain AC-CE-EB-BD-DA to the first row of Table 4. Either set of 15 bases provides a parity proof of the BKS theorem. There are a total of 120 complementary pairs of 30-ray critical sets of the type shown here.

## PATTERNS OF 36-RAY CRITICAL SETS

AC-AD-AE-BC-BD-BE
AB-AD-AE-CB-CD-CE
AB-AC-AE-DB-DC-DE
AB-AC-AD-EB-EC-ED
BA-BD-BE-CA-CD-CE
BA-BC-BE-DA-DC-DE
BA-BC-BD-EA-EC-ED
CA-CB-CE-DA-DB-DE
CA-CB-CD-EA-EB-ED
DA-DB-DC-EA-EB-EC

TABLE 7. A 36-ray critical set can be obtained by keeping all the rays in any row of Table 4 belonging to one of the chains of dual line pairs shown above. The procedure for constructing these sets is identical to that for the 30-ray sets explained in the caption to Table 5. If all the letters are primed and the chains are applied to the columns rather than the rows, the remaining critical sets are obtained.

# A 36-RAY CRITICAL SET

|     | A | B | C | D | E |
|-----|---|---|---|---|---|
| A'  | **1** **2** 3 **4**<br>**5** **6** **7** 8<br>9 10 **11** **12** | **31** **42** **51** 16<br>**38** 24 **58** **25**<br>**56** **45** 17 **35** | **22** 60 39 **28**<br>18 **47** 33 55<br>13 32 **50** 41 | **57** **23** 27 40<br>**36** 53 **20** 46<br>**43** 49 30 **14** | 44 **29** **15** 52<br>59 26 **37** **21**<br>34 **19** 48 **54** |
| B'  | **13** **14** **15** 16<br>17 **18** **19** **20**<br>**21** **22** **23** 24 | **43** **54** 3 **28**<br>**50** **36** 10 **37**<br>8 **57** **29** **47** | 34 **12** **51** 40<br>30 59 **45** **7**<br>**25** 44 **2** 53 | 9 **35** 39 52<br>48 **5** 32 **58**<br>55 **1** **42** 26 | **56** 41 27 **4**<br>**11** **38** 49 33<br>46 **31** 60 **6** |
| C'  | **25** 26 27 **28**<br>**29** 30 **31** 32<br>33 34 **35** **36** | 55 **6** **15** 40<br>2 48 **22** 49<br>**20** 9 41 59 | 46 24 3 52<br>**42** **11** **57** **19**<br>**37** **56** **14** **5** | **21** **47** **51** **4**<br>60 17 44 10<br>7 **13** **54** **38** | 8 53 39 16<br>**23** **50** **1** **45**<br>**58** **43** **12** **18** |
| D'  | **37** **38** 39 40<br>41 **42** **43** 44<br>**45** 46 **47** 48 | 7 **18** 27 **52**<br>**14** 60 34 **1**<br>32 **21** 53 **11** | **58** **36** **15** **4**<br>**54** **23** **9** **31**<br>49 8 26 17 | 33 59 3 16<br>**12** **29** **56** **22**<br>**19** **25** **6** **50** | **20** **5** **51** **28**<br>**35** **2** **13** **57**<br>10 55 24 30 |
| E'  | 49 **50** **51** 52<br>53 **54** 55 **56**<br>**57** **58** 59 60 | **19** 30 39 **4**<br>26 **12** 46 **13**<br>44 33 **5** **23** | 10 48 27 16<br>**6** **35** **21** **43**<br>**1** **20** **38** **29** | **45** **11** **15** **28**<br>24 41 8 34<br>**31** **37** **18** **2** | 32 17 3 40<br>**47** **14** **25** **9**<br>**22** **7** **36** **42** |

TABLE 8. Shown in bold are the 36 rays picked out by the line chain AC-AD-AE-BC-BD-BE of Table 7 applied to the first row of Table 4. These rays make up 18 "intact" bases, 12 "almost intact" bases (containing a single deleted ray) and 36 "half intact" bases (containing two deleted rays).

# BKS PROOF BASED ON A 36-RAY SET

| Green Rays | | | | | | | All-Red Basis | | | |
|---|---|---|---|---|---|---|---|---|---|---|
| 2 | 7  | 10 | <u>19</u> | 53 |    |    | 3  | 24 | 46 | 52 |
| 2 | 7  | 16 | <u>58</u> | 21 | 46 | 19 | 8  | 17 | 26 | 49 |
| 2 | 36 | 24 | 18 | <u>6</u>  | 11 |    | 3  | 17 | 32 | 40 |
| 2 | 36 | 24 | 27 | <u>6</u>  | 39 | 12 | 11 | 19 | 42 | 57 |
| 2 | 36 | 52 | 8  | <u>12</u> |    |    | 10 | 16 | 27 | 48 |
| 2 | 36 | 52 | 16 | 30 | <u>12</u> |    | 11 | 19 | 42 | 57 |
| 2 | 36 | 52 | 16 | 55 | <u>19</u> |    | 8  | 17 | 26 | 49 |
| 2 | 42 | 30 | 12 | <u>52</u> |    |    | 4  | 15 | 36 | 58 |
| 2 | 42 | 30 | 46 | 8  | <u>21</u> | 40 | 10 | 16 | 27 | 48 |
| 2 | 42 | 30 | 46 | 16 | <u>18</u> |    | 4  | 15 | 36 | 58 |
| 2 | 42 | 39 | 10 | <u>18</u> |    |    | 8  | 17 | 26 | 49 |
| 2 | 42 | 39 | 24 | <u>12</u> | 15 |    | 6  | 21 | 35 | 43 |

TABLE 9. "Proof-tree" demonstrating the impossibility of satisfactorily coloring the 36-ray set shown in bold in Table 8. The proof uses only the 18 "intact" and 12 "nearly intact" bases of Table 8. The proof begins by looking at the "intact" basis 2 48 22 49. One ray in it must be colored green, and we can choose it to be 2 without loss of generality because the other rays can be rotated into 2, by suitable symmetry operations of the polytope, while maintaining the invariance of the 36-ray set as a whole. Coloring 2 green forces 22 to be red. The basis 22 7 36 42 must then have one of the rays 7, 36 or 42 colored green. The different branches of the proof-tree show what happens if one makes different choices for the green rays at various points of the coloring process. Unforced green rays are not underlined, but forced ones are. The branching of the proof-tree stops when green rays become forced, and every terminal branch leads eventually to a basis in which all the rays are colored red (this basis is shown in bold at the end of the branch). Thus no viable coloring of these rays is possible and the BKS theorem is proved.

### CRITICAL SETS OF RAYS IN FOUR DIMENSIONS

| Rays | Critical Set size | # of critical sets | References |
|---|---|---|---|
| 24 rays of Peres | 18 | 16 | [5] |
|  | 20 | 96 | [6] |
| 40 rays of Zimba-Penrose | 28 | 5 | [8] |
|  | 32 | 45 | [16] |
| 60 rays of the 600-cell | 30 | 240 | Present work |
|  | 36 | 200 |  |

TABLE 10. The smallest known critical sets of rays in four dimensions.